\documentclass[12pt,preprint]{aastex}
\usepackage{natbib}
\usepackage{rotating}
\usepackage{mathptmx}

\newcommand{\ionhy}{H{\sc ii}}
\newcommand{\UCHII}{UCH{\sc ii}}
\newcommand{\kms}{$\mbox{km~s}^{-1}$}
\newcommand{\cc}{$\mbox{cm}^{-3}$}
\newcommand{\cs}{$\mbox{cm}^{-2}$}

\newcommand{\twospec}[2] {
 \begin{center}
    \begin{minipage}[t]{0.45\textwidth}
      \includegraphics[angle=270,scale=0.4]{#1}
    \end{minipage}
    \hfill
    \begin{minipage}[t]{0.45\textwidth}
      \includegraphics[angle=270,scale=0.4]{#2}
    \end{minipage}
  \end{center}
}

\shorttitle{Two New Class II Methanol Maser Transitions}
\shortauthors{S. P. Ellingsen et al.}

\begin{document}

\title{Discovery of Two New Class II Methanol Maser Transitions in G\,345.01+1.79}

\author{S. P. Ellingsen}
\affil{School of Mathematics and Physics, University of Tasmania, Private Bag 37, Hobart, TAS 7001, Australia}
\email{Simon.Ellingsen@utas.edu.au}
\author{A. M. Sobolev}
\affil{Ural Federal University, Lenin ave. 51, 620000 Ekaterinburg, Russia}
\author{D.M. Cragg, P.D. Godfrey}
\affil{School of Chemistry, Building 23, Monash University, Victoria 3800, Australia}

\begin{abstract}
We have used the Swedish ESO Submillimetre Telescope (SEST) to search for new class II methanol maser transitions towards the southern source G\,345.01+1.79.  Over a period of 5 days we observed 11 known or predicted class II methanol maser transitions.  Emission with the narrow line width and characteristic velocity of class~II methanol masers (in this source) was detected in 8 of these transitions, two of which have not previously been reported as masers. The new class II methanol maser transitions are the $13_{-3}-12_{-4}\,E$ transition at 104.1~GHz and the $5_1-4_2\,E$ transition at 216.9~GHz.  Both of these are from transition series for which there are no previous known class II methanol maser transitions.  This takes the total number of known class II methanol maser series to 10, and the total number of transitions (or transition groups) to 18.  The observed 104.1~GHz maser suggests the presence of two or more regions of masing gas with similar line of sight velocities, but quite different physical conditions.  Although these newly discovered transitions are likely to be relatively rare, where they are observed combined studies using the Australia Telescope Compact Array and the Atacama Large Millimeter Array offer the prospect to be able to undertake multi-transition methanol maser studies with unprecedented detail.

\end{abstract}


\section{Introduction}

Class~II methanol masers are unique tracers of the early stages of high-mass star formation \citep[see e.g.][]{Ellingsen06}.  The other maser transitions commonly observed in high-mass star formation regions (main-line OH and 22 GHz water) are also observed in different environments, such as late-type stars and low-mass star formation regions.  In contrast, despite a number of sensitive and extensive searches no class~II methanol masers have been found associated with anything other than high-mass star formation regions \citep{Ellingsen+96,Minier+03,Xu+08}.  The majority of class~II methanol maser sources are not associated with ultra-compact \ionhy\  (\UCHII) regions \citep{Phillips+98,Walsh+98}.  This indicates that either the masers are associated with objects that are not massive enough to form detectable \ionhy\ regions, or that they are associated with high-mass stars prior to the formation of an \UCHII\ region.  Numerous observations at millimeter through mid-infrared wavelengths \citep[e.g.][]{Walsh+03,Ellingsen06,Cyganowski+09} suggest that many class~II methanol maser sources are associated with a very early evolutionary phase in the high-mass star formation process.

Hindered rotation in the methanol molecule leads to a very large number number of rotational and vibrational transitions in the millimetre and submillimetre regime.  A large number of these transitions have been observed in interstellar space and a select few show maser action.  The transitions in which masers are observed have been empirically classified into two groups \citep{Batrla+87,Menten91}.  The class~II methanol masers are closely associated with sites of high-mass star formation and related objects, including infrared sources, \ionhy\ regions and often OH or water masers.  The 6.7 GHz $5_1 - 6_0\mbox{A}^{+}$ transition is the strongest and most common class~II maser transition, having been observed towards more than 900 regions within our Galaxy \citep*{Caswell+10,Green+10,Caswell+11}.  Class~I methanol masers are found near high-mass star formation regions, but offset from the \ionhy\ regions by a few tenths of a parsec \citep*{Kurtz+04}.  The 44.0~GHz $7_0 - 6_1\mbox{A}^{+}$ transition is the strongest and most common class~I maser transition.  This paper focuses solely on class~II methanol masers and hereafter it is implicit that all references to ``methanol masers'' refer to class~II masers.

To date eighteen different transitions (or transition groups), associated with eight different methanol transition series have been observed as class~II methanol masers and their frequencies, along with the discovery reference are listed in Table~\ref{tab:classII}.  Counting the total number of known class~II methanol maser transitions is somewhat subjective.  The distinction which we have made is that where there are two or more transitions at similar frequencies which are always (or nearly always) either present, or absent together from sources we have considered these to be a transition group.  For example, all observations to date of the 38.3 ($6_{2} -5_{3}\mbox{~A}^{-}$) and 38.5~GHz ($6_{2} -5_{3}\mbox{~A}^{+}$) transitions show that where emission from one of these is detected, the other is also and with similar intensity \citep{Haschick+89,Ellingsen+11}.  Hence we consider the 38.3 and 38.5~GHz masers to represent a single transition group.    In contrast the 6.7, 107.0 and 156.6~GHz transitions are all members of the $J_{1} - (J+1)_{0}$A$^+$ series, however, many sources which show emission in the 6.7~GHz transition are not observed in either of the other two, so we count these as three separated maser transitions.  The $J_{0} - J_{-1}$E series at frequencies near 157~GHz, is another example of a single transition group \citep{Slysh+95}, although the 148.1~GHz $15_{0} - 15_{-1}$E is counted as a separate transition as it stands apart both in frequency, and in the physical conditions it requires.

The two lowest frequency methanol maser transitions ($5_1 - 6_0\mbox{A}^{+}$ at 6.7~GHz and $2_0 - 3_{-1}\mbox{E}$ at 12.2~GHz) have been detected towards a large number of sources, but the majority of the other transitions are only observed towards a handful of sources \citep[see][and references therein]{Ellingsen+11}. Current models of class~II methanol masers \citep{Sobolev+97a,Cragg+02,Cragg+05} find that the 6.7 and 12.2~GHz transitions are strongly inverted over a wide range of physical conditions, explaining both their high-brightness temperatures and common occurrence.  The models successfully predict masing in all of the other class~II transitions for which it has been observed, but generally the physical conditions implied are much more restricted \citep{Sobolev+97b}. So the presence of multiple class~II maser transitions within one source can, through comparison with models, be used to constrain the physical conditions in the masing region (under the assumption that the various transitions all arise from the same location).  

\section{Observations}

The observations were made using the Swedish ESO Submillimetre Telescope (SEST) between 2000 June 22-27.  We observed a total of 13 different methanol transitions, 7 of which are known class~II methanol maser transitions and the remaining 6 of which are predicted class II methanol maser transitions \citep{Sobolev+97b}.  The frequencies of the observed transitions range from 84 to 265 GHz.  These transitions fall into 11 different transition groups, as the $6_{2} -5_{3}\mbox{A}^{-}/\mbox{A}^{+}$ and $10_{2} -9_{3}\mbox{A}^{-}/\mbox{A}^{+}$ represent a single transition group in each case.  All of the observations were targeted  towards the class II methanol maser source G\,345.01+1.79 (more precisely G\,345.010+1.792) at $\alpha_{J2000}$ = 16$^h$56$^m$47.58$^s$ ; $\delta_{J2000}$ = -40$^{\circ}$14$^{\prime}$25.8$^{\prime\prime}$ \citep{Caswell09}. The onsource integration time for each of these transitions ranged from 30 to 60 minutes and the resulting RMS per spectral channel range from 0.3 to 2.3 Jy.  The observations were made by frequency switching with a throw of 6~MHz and a switching time of 2 minutes.  The spectra were obtained using the high-resolution acousto-optic spectrometer, which has a total bandwidth of 86 MHz and a spectral resolution of 80~kHz, resulting in velocities resolutions of between 0.28 and 0.09~\kms\/ (for the lowest and highest frequency transitions observed, respectively).  The intensity scale was converted from Kelvin to Jy by assuming at sensitivity of 25 Jy K$^{-1}$ for the transitions at frequencies less than 150 GHz, 30 Jy K$^{-1}$ for transitions at frequencies between 150 and 200 GHz and 41 Jy K$^{-1}$ for transitions at frequencies higher than 200 GHz.

The data were processed using Per Bergman's {\tt XS} package, following the standard procedure for radio-millimetre single dish spectral line observations.  The details of the transitions observed and the achieved sensitivities are summarized in Table~\ref{tab:obssum}.  Because we were searching for narrow maser lines the spectra were not smoothed prior to measuring the RMS.

\section{Results} \label{sec:results}

\begin{figure*}
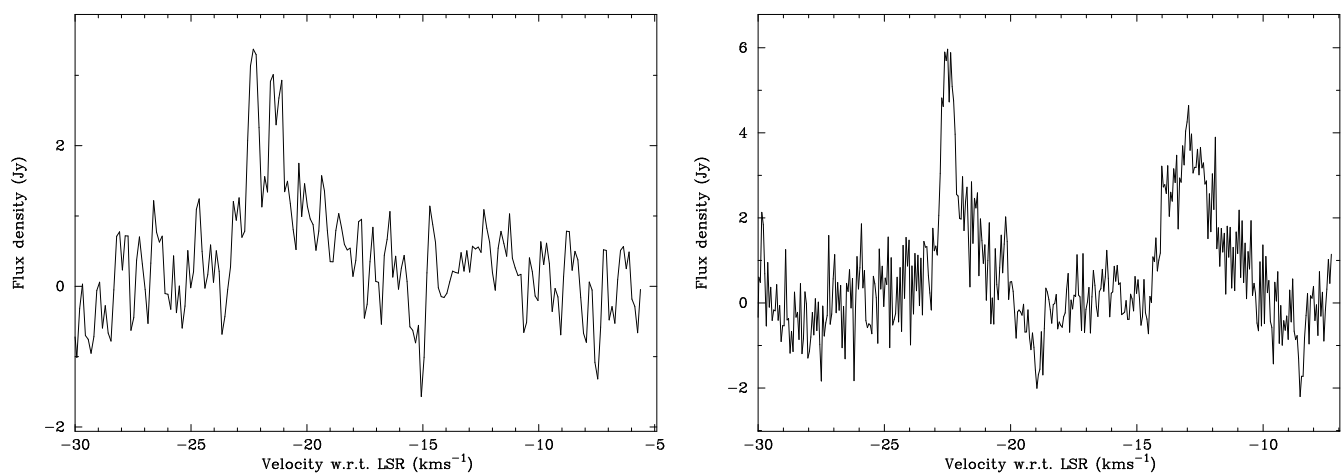

\twospec{g345_104}{g345_217}
\caption{Spectra of the 104.1 GHz ($13_{-3}-12_{-4}\,E$) (left) and 216.9 GHz ($5_1-4_2\,E$) (right) transitions of methanol towards G\,345.01+1.79} \label{fig:new}
\end{figure*}

\begin{figure}
\includegraphics[scale=0.5]{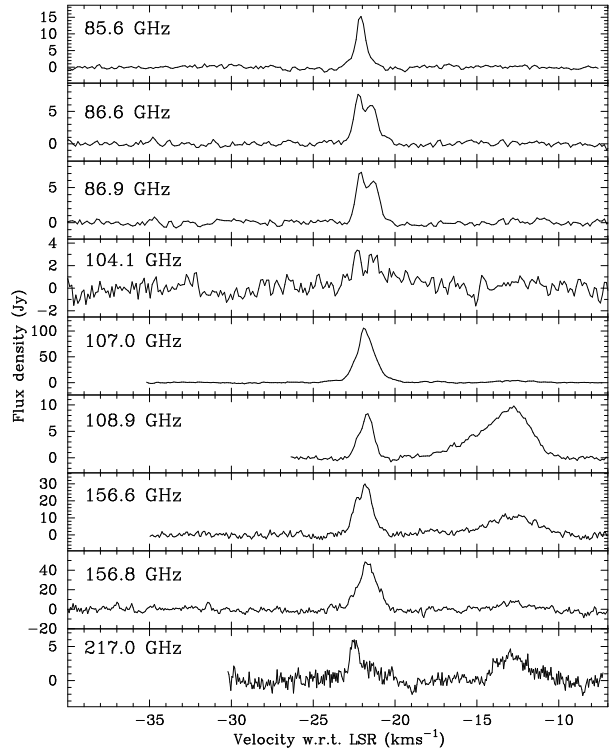}
\caption{Spectra of the detected class II methanol maser transitions towards G\,345.01+1.79} \label{fig:stack}
\end{figure}

Emission was detected at 9 of the 13 methanol transitions observed.  These observations represent the first astronomical detection of maser emission from the 104.1 GHz ($13_{-3}-12_{-4}\,E$) transition and the 216.9 GHz ($5_1-4_2\,E$) transition.   The spectra of all of the methanol maser transitions detected during the current SEST observations are shown in Fig.~\ref{fig:stack}.  It should be noted that the class~II methanol maser reported here is a different transition from the 104.3~GHz ($11_{-1}-10_{-2}\,E$) class~I methanol maser previously discovered by \citep{Voronkov+05}.  Spectra of the emission in these two new class II methanol maser transitions are shown in Fig~\ref{fig:new}.  In both cases the maser emission consists of a narrow (FWHM $<$ 0.5~\kms) peak at a velocity of around -22~\kms\/.  This is the approximate velocity at which most of the rare class II methanol masers observed towards G\,345.01+1.79 peak.  Observations of the 104.1 GHz transition of methanol have previously been made towards the archetypal class~II methanol maser region W3(OH) by \citet{Sutton+04}.  No emission was detected in the W3(OH)/(H$_2$O) region at the sensitivity level achieved for their 104.1~GHz observations.

The 216.9 GHz spectrum shows an additional broader peak at approximately -13~\kms, coinciding with the velocity at which thermal methanol emission has previously been observed in this source \citep{Valtts+99,Salii+06}. Although the signal to noise in each of these spectra is moderately low, subsequent SEST observations focusing on studying the thermal emission in this region have confirmed the presence of the maser emission in each of these two transitions \citep{Salii+06}.  The observations of the 104.1 and 216.9~GHz transitions of \citet{Salii+06}, were made in April 2003, slightly less than three years after the observations reported here.  In 2003 the 104.1~GHz maser had a peak flux density of $\sim$ 3.8~Jy, consistent (within the relative uncertainties) with the flux density of the current SEST observations.  In 2003 the 216.9~GHz transition had a peak flux density of $\sim$ 10.3~Jy, significantly greater than the $\sim$ 6~Jy reported in the current observations.  Comparing the 216.9~GHz spectrum in figure~1 in \citet{Salii+06} with our Figure~\ref{fig:new} shows that relative intensity of the maser and thermal components has not change dramatically.  Given that we don't expect to see any significant variations in the thermal emission, the observed difference in maser intensity is likely primarily due to calibration and/or pointing errors in the current observations, rather than significant variations in the maser.

\citet{Salii+06} detected maser emission from the 231.3~GHz ($10_{2}-9_{3}\mbox{A}^-$) transition in G\,345.01+1.79 with a peak flux density of $\sim$2~Jy.  No maser emission was detected in our SEST observations, however, as they are significantly less sensitive (RMS 1.5~Jy) than the observations of \citet{Salii+06}, this is expected.

We have fitted Gaussian profiles to both the maser emission, and (where seen) the thermal emission for each transition.  The results of these fits are summarised in Table~\ref{tab:gauss}.  For all of the transitions it was necessary to fit more than one Gaussian component to adequately describe the maser emission (i.e. residuals comparable with the noise).  Many of the transitions have the strongest components at velocities close to -22.2 and -21.4~\kms\/ (e.g. 86.6, 86.9 and 104.1 GHz), while others show their strongest emission at a velocity of around -21.7~\kms\/ (e.g. 107.0, 108.9, 156.6 and 156.8 GHz).  Even accounting for the relatively low signal to noise in some of the spectra it is clear from this figure that although the maser emission in all transitions peaks at a velocity of around -22~\kms, it cannot be described by a single Gaussian component.  Uncertainty in the absolute frequency calibration of the spectrometer and in the adopted rest frequencies of some of the transitions is approximately 0.2~\kms, nevertheless, the difference in the shape and width of the emission at different frequencies requires the presence of two or more different maser components. 

\section{Discussion}

G\,345.01+1.79 has been observed at 14 of the 15 class~II methanol maser transition (or transition groups) frequencies listed in Table~\ref{tab:classII}.  The exception is the $8_{2}-9_{1}\,A^{-}$ (29.0~GHz) transition, for which there is only one published study \citep{Wilson+93}.  Including the two new class~II transitions reported here, maser emission has been detected in G\,345.01+1.79 for 15 of the 16 known class~II transitions (or transition groups) observed in this source, the exception being the $9_{2}-10_{1}\,A^{+}$ 23.1~GHz, for which the sensitive observations of \citet{Cragg+04} placed a 3$\sigma$ upper limit of 0.5 Jy.  Some other well studied \ionhy\/ regions (such as W3(OH) and NGC6334F) also show a large number of class II methanol masers, but no others exhibit as many as G\,345.01+1.79.  It is not clear why G\,345.01+1.79 is such an exceptional source.  \citet{Norris+93} identified two distinct class II methanol maser sites (separated by 19$\arcsec$, with non-overlapping velocity ranges) in this general region.  G\,345.010+1.792 has methanol emission covering the velocity range -24 to -16~\kms, while G\,345.012+1.797 masers covers -16 to -10 \kms\/ \citep{Caswell09}.  G\,345.010+1.792 is associated with an UCH{\sc ii} region, and all of the rare/weak methanol maser transitions are associated with this source.  \citet{Breen+10} detected three water masers in this region, one associated with each maser site and a third which is not associated with any other known maser transitions.  Interestingly, the weakest water maser in the region is associated with the strongest methanol maser site (G\,345.010+1.792), which also has an associated OH maser \citep{Caswell98}, while the secondary methanol maser site is associated with the strongest water maser.  In this paper, we are focusing purely on the strongest methanol maser site (G\,345.010+1.792).  

In general the strongest emission from class II methanol maser transitions in this source is at velocities near -22~\kms.  In contrast, when it was first observed in the early 1990's, the peak velocity of the 6.7 GHz methanol maser was at -17~\kms\/ \citep{Caswell+95a}.  The intensity of that spectral feature has declined over the intervening 20 years \citep[see e.g.][]{Ellingsen+04,Caswell+10}, and the peak of the 6.7~GHz maser emission is now at $\sim$ -21~\kms.  The 19.9~GHz methanol maser is now the only transition which peaks at -17~\kms\/ \citep{Ellingsen+04}, and in contrast to the other rare/weak methanol maser transitions no emission is observed at velocities less than -19~\kms.

The complex, non-linear relationship between the physical conditions of the masing gas and the intensity of the observed maser means that it is generally not possible to uniquely infer the former from the latter.  Where more than one maser transition is observed from the same volume of gas, it must be the case that a complete theoretical maser model will be able to find conditions that simultaneously invert all of the detected transitions in the observed intensity ratios.    \citet{Cragg+01} used information on 9 class~II methanol maser transitions, collected with a variety of instruments, over a period of several years, to constrain the physical conditions in G\,345.01+1.79, using the model of \citet{Sobolev+97a}.  They found the observed ratio of maser intensities to be consistent with cool gas (30K) at a density of $10^6$~\cc\, with a methanol column density of $5 \times 10^{17}$~\cs, being pumping by the radiation from warm (175K) dust.

Modelling such as that undertaken by \citet{Cragg+01} requires a number of assumptions.  In particular, it is assumed that the emission from each transition arises from the same region of gas and that the emitting region is the same size for each transition.  Where information is collated from observations made with different instruments over a period of years, it is also necessary to ignore the effects of temporal variability.  For G345.01+1.79 \citet{Minier04} showed that the 6.7, 12.2, 85.5 and 86.6 GHz transitions are all coincident to within 1$\arcsec$ (the relative positional accuracy of the observations), and where it has been possible to test it the assumption that the different maser transitions are co-spatial with greater precision this appears to be generally true \citep{Menten+92,Norris+93,Sutton+01}.   The assumption that the emitting regions are the same size in each transition is not likely to be correct, but it requires very long baseline interferometry observations to make the necessary measurements, and at present that is only practical for the 6.7 and 12.2 GHz transitions.  Although the variability of class~II methanol masers is not as extreme as that observed for water masers, variations on timescales of months or years are common \citep{Caswell+95b,Goedhart+04}.  For the weaker millimetre methanol masers variability is also expected, possibly even to a greater extent due to the lower degree of saturation in these transitions, and evidence of variations in the millimetre masers in G\,345.01+1.79 have been observed \citep{Ellingsen+03}.  

Of the many thousands of methanol transitions which fall at centimeter or millimeter wavelengths, a small subset (around 20) exhibit pronounced maser characteristics.  Many of the maser lines can be grouped into series with the same patterns of $J$ and $K$ quantum numbers.  For example, the $5_1 - 6_0\mbox{A}^{+}$ maser at 6.7 GHz is part of a series of $J_{1} - (J+1)_{0}$A$^+$ masers including the $3_1 - 4_0\mbox{A}^{+}$ maser at 107.0 GHz and the $2_1 - 3_0\mbox{A}^{+}$ maser at 156.6 GHz. This is consistent with the models, which also predict maser action in the $4_1 - 5_0\mbox{A}^{+}$ transition at 57.0 GHz (which cannot be observed by ground-based telescopes due to atmospheric absorption by molecular oxygen at this frequency).  One can think of the $K=1$ ladder of energy levels as being slightly overpopulated with respect to the nearby $K=0$ energy levels, with the resulting population inversions being responsible for the maser action.  The new maser transitions detected in the current work are the first members of two new series of methanol transitions to exhibit maser action, and as such represent a more stringent test of the models.   Our detection of maser action in the $5_1-4_2\,E$  transition at 216.9 GHz is consistent with the modelling of \citet{Sobolev+97b} and \citet{Cragg+05}, which also include the $3_1-2_2\,E$  120.2 GHz, $4_1-3_2\,E$ 168.6, and $6_1-5_2\,E$  265.3 GHz transitions from the same series in the list of maser candidates.  The newly observed $13_{-3}-12_{-4}\,E$ maser at 104.1 GHz is part of the same series as the $11_{-3}-10_{-4}\,E$ 7.3 GHz and $12_{-3}-11_{-4}\,E$ 55.7 GHz transitions, both of which were identified as class II methanol maser candidates in these models.  In the models presented in these papers the 104.1 GHz maser candidate was below the threshold intensity for listing, indicating that its excitation to observable levels requires more extreme conditions.

Detailed modelling of the methanol masers in G345.01+1.79 is beyond the scope of this letter, and will be the subject of a future publication.  In addition to the data presented here, previous and subsequent observations with SEST and other instruments have obtained detections and limits on the emission from more than three dozen known or potential class II methanol transitions.  Here we restrict ourselves to a preliminary exploration which utilises the data from these observations, in combination with published 6.7 and 12.2 GHz spectra, following the approach outlined in \citet{Cragg+01}.  The significant differences in the detailed spectral structure of the methanol maser emission in different transitions near velocities of -22~\kms, demonstrates the presence of multiple, partially blended maser components.  This implies that within G345.01+1.79 there are multiple sites with differing physical conditions which are giving rise to emission in rare methanol transitions.  Hence it is unlikely that any single model fit will be able to explain the observed intensities of all of the transitions, and this is indeed the case.

To adequately explain the observed (and the one non-detected) class~II methanol transitions in G345.01+1.79 using the \citet{Sobolev+97a} model we require emission from at least two regions with similar/overlapping line-of-sight velocities but different physical parameters.  This is inferred from the differences in the line profiles of the observed transitions (e.g. Figure~\ref{fig:stack} shows the 86.6, 86.9 and 104.1 GHz transitions are double-peaked, while other transitions are single-peaked) and in their peak velocities (see Table~\ref{tab:gauss}).  Applying models which assume that emission in all the transitions is co-spatial cannot produce a single, self-consistent solution in cases such as this.  For example in G345.01+1.79 we observe both maser transitions which require low densities (e.g. the 86.6 and 86.9~GHz) and transitions which prefer higher densities (e.g. the 85.6 and 104.1~GHz) for their efficient excitation.  We suggest that the reason for G345.01+1.79 exhibiting maser emission in an unusually large number of different methanol transitions, may be that the coexistence of masing regions covering such a broad range of physical conditions is rare.

\section{Conclusions}

We have discovered two, new class~II methanol maser transitions towards the G345.01+1.79 star formation region.  The emission from these transitions is relatively weak in G345.01+1.79 and it is likely that a only small number of high-mass star formation regions will show maser emission in these transitions. G345.01+1.79 clearly has great potential for helping us to study the physical conditions in high-mass star formation regions at high resolution, through observations of the large number of class~II methanol maser transitions.  The primary limitations on the current observations are due to relatively low signal to noise and spatial blending.  High resolution observations of the weaker millimetre methanol masers to obtain detailed information on their distribution, to constrain the spot size and determine whether the different transitions are coincident have recently become much more feasible with the commencement of ALMA operations.  Such observations offer the prospect of unlocking the potential of multi-transition maser observations and providing important insights into the processes which occur in high-mass star formation regions.

\section*{Acknowledgements}

This research has made use of NASA's Astrophysics Data System Abstract Service.  Travel to SEST for observations was provided under the access to major research facilities program which is supported by the Commonwealth of Australia under the International Science Linkages program.  AMS was partly supported by Russian state contract No. 14.518.11.7064 and the Russian Foundation for Basic Research (grants 10-02-00589-a and 11-02-01332-a).  DMC and PDG acknowledge financial support for this work from the Australian Research Council.

\begin{table}
  \begin{center}
  \caption{Class~II methanol maser transitions.  The rest frequencies have 
    been rounded at the 100~kHz significance level.  The G\,345.01+1.79 reference
    is the most sensitive high resolution observation available for the specific transition.}
  \begin{tabular}{clcll} \hline
  {\bf Rest}      & {\bf Transition} & {\bf Series} &  {\bf Discovery} & {\bf G\,345.01+1.79} \\ 
  {\bf Frequency} &                      &                      & {\bf Reference}    & {\bf Reference} \\  \hline 
  6.7   & $5_1 - 6_0\mbox{A}^{+}$            & 1 & \citet{Menten91b}  & \citet{Norris+93} \\  
  12.2  & $2_0 - 3_{-1}\mbox{E}$              & 2 & \citet{Batrla+87}     & \citet{Norris+88} \\        
  20.0  & $2_{1} - 3_{0}\mbox{E}$          & 3 & \citet{Wilson+85}       & \citet{Ellingsen+04} \\  
  23.1  & $9_{2} - 10_{1}\mbox{A}^{+}$ & 4 & \citet{Wilson+84}   & \citet{Cragg+04} \\      
  29.0  & $8_{2} - 9_{1}\mbox{A}^{-}$    & 4 & \citet{Wilson+93}     &                                \\       
  37.7  & $7_{-2} - 8_{-1}\mbox{E}$       & 5 & \citet{Haschick+89}  & \citet{Ellingsen+11} \\  
  38.3/38.5  & $6_{2} -5_{3}\mbox{A}^{-}/\mbox{A}^{+}$    & 6 & \citet{Haschick+89} & \citet{Ellingsen+11} \\ 
  85.6  & $6_{-2} -7_{-1}\mbox{E}$        & 5 & \citet{Cragg+01}         & \citet{Cragg+01}   \\ 
  86.6/86.9  & $7_{2} - 6_{3}\mbox{A}^{-}/\mbox{A}^{+}$   & 6 & \citet{Cragg+01}      & \citet{Cragg+01} \\      
  107.0 & $3_{1} - 4_{0}\mbox{A}^{+}$  & 1 & \citet{Valtts+95}      & \citet{Caswell+00} \\      
  108.9 & $0_{0} - 1_{-1}\mbox{E}$        & 2 & \citet{Valtts+99}         & \citet{Valtts+99} \\         
  148.1 & $ 15_{0} - 15_{-1}\mbox{E}$   & 7 & \citet{Salii+06}       & \citet{Salii+06} \\ 
  156.6 & $2_{1} - 3_{0}\mbox{A}^{+}$  & 1 & \citet{Slysh+95}      & \citet{Caswell+00} \\       
  157.3 & $J_{0} - J_{-1}\mbox{E}$ group & 7 & \citet{Slysh+95} & \citet{Slysh+95}  \\ 
  165.0 & $J_{1} - J_{0}\mbox{E}$ group  & 8 & \citet{Salii+03}    & \citet{Salii+03} \\ 
  231.3 & $10_2-9_3\mbox{A}^-$              & 6  & \citet{Salii+06}    & \cite{Salii+06} \\ \hline
  \end{tabular}
  \label{tab:classII}
  \end{center}
\end{table}

\begin{table*}
  \begin{center}
  \caption{Summary of lines observed towards G\,345.01+1.79.  The rest frequencies for transitions $<$ 200~GHz are from \citet{Tsunekawa+95}, the remainder are from \citet{Sastry+84}}
  \begin{tabular}{lrrrrrr} \hline
{\bf Transition} & {\bf $\nu$} & {\bf RMS} & {\bf T$_{sys}$} & {\bf Integration} & {\bf Peak Flux} & {\bf Velocity} \\
                           & {\bf (MHz)} &{\bf (Jy)} & {\bf (K)}  & {\bf (minutes)}                  & {\bf (Jy)}            & {\bf \kms}        \\ \hline
 $13_{-3}-14_{-2}\,E$ &  84423.769   & 0.4 &  143 & 50   & $<$1.9 &                \\
 $6_{-2}-7_{-1}\,E$     &  85568.084    & 0.4 &  160 & 54   & 10.9(1.9)  & -22.1 \\
 $7_2-6_3\,A^-$          &  86615.578    & 0.3 &  149 & 42   & 7.2(0.6)    & -22.3 \\
 $7_2-6_3\,A^+$         &  86902.956    & 0.3 &  140 & 60   & 6.8(0.5)    & -22.1 \\
 $13_{-3}-12_{-4}\,E$ & 104060.647  & 0.7 &  254 & 30   & 2.8(1.1)    & -22.3 \\
 $3_{1}-4_{0}\,A^+$    & 107013.812  & 0.7 &  265 & 56   & 66(8)        & -21.9  \\
 $0_0-1_{-1}\,E$          & 108893.948  & 0.3 &  166 & 60   & 8.3(0.4)    & -21.7 \\
 $2_1-3_0\,A^{+}$       & 156602.413  & 1.3 &  600 & 56   & 27.6(1.2) & -21.8  \\
 $7_0-7_{-1}\,E$          & 156828.533  & 2.3 & 1015 & 54  & 45.0(1.8) & -21.7  \\
 $5_1-4_2\,E$              & 216945.600  & 0.8 &  203 & 60    & 4.2(0.8)  & -22.5   \\
 $10_2-9_3\,A^-$         & 231281.100  & 1.5 &  267 & 42   & $<$ 7.5 &        \\
 $10_2-9_3\,A^+$        & 232418.590  & 1.6 &  329 & 30   & $<$ 7.8 &        \\
 $4_2-5_1\,A^-$           & 234683.390  & 1.3 &  315 & 50   & $<$ 6.4 &       \\
\hline
\end{tabular}
\label{tab:obssum}
\end{center}
\end{table*}

\begin{table}
  \begin{center}
  \caption{Gaussian fitting of methanol maser transitions in G\,345.01+1.79, with formal
   uncertainty from fitting in parenthesis.}
  \begin{tabular}{rrrr} \hline
{\bf Transition} & {\bf Peak} & {\bf Velocity} & {\bf FWHM} \\
{\bf (MHz)}        & {\bf (Jy)}    & {\bf (\kms)}    & {\bf (\kms)}  \\ \hline
 85.6  & 10.9(1.9) & -22.10(0.01) & 0.51(0.07) \\
           & 4.9(2.0)   & -22.06(0.06) & 1.25(0.23) \\
 86.6  & 7.2(0.6)   & -22.25(0.05) & 0.66(0.09) \\
           & 6.0(0.5)   & -21.41(0.04) & 0.77(0.18) \\
           & 1.0(0.5)   & -20.48(0.22) & 0.52(0.43) \\
 86.9  & 6.8(0.5)   & -22.14(0.03) & 0.60(0.06) \\
           & 6.0(0.4)   & -21.32(0.03) & 0.82(0.10) \\
104.1 & 2.8(1.1)   & -22.31(0.08) & 0.44(0.21) \\
            & 1.9(1.0)  & -21.35(0.16) & 0.64(0.41) \\
            & 1.1(0.5)  & -20.66(0.90) & 5.0(1.9)  \\
107.0 & 66(8)       & -21.85(0.12) & 0.98(0.14) \\
           & 29(8)       & -21.63(0.15) & 1.73(0.22) \\
           & 6(4)         & -22.14.(0.42) & 3.11(0.52) \\
           & 2.0(0.4)  & -17.65(0.23)  & 1.78(0.50) \\
           & 3.7(0.3)  & -12.65(0.11)  & 2.55(0.27) \\
108.9 & 2.3(0.8)  & -22.36(0.16)  & 0.47(0.17) \\
           & 8.3(0.4)  & -21.67(0.14)  & 0.82(0.08) \\
           & 4.0(1.3)  & -14.65(0.94)  & 3.80(1.16) \\
           & 7.3(2.9)  & -12.62(0.14)  & 2.43(0.33) \\
156.6 & 10.5(3.6) & -22.49(0.06) & 0.48(0.15) \\
           & 27.6(1.2) & -21.79(0.04) & 0.95(0.11) \\
           & 10.5(0.6) & -12.80(0.22) & 3.31(0.24) \\
156.8 & 45(1.8)   & -21.68(0.03) & 1.36(0.06) \\
           & 6.6(0.6)  & -12.80(0.22) & 1.99(0.51) \\
216.9 & 3.6(0.3) & -12.82(0.10)  & 2.31(0.23) \\
           & 4.2(0.8) & -22.50(0.10)  & 0.49(0.11) \\
           & 2.0(0.5) & -22.02(0.23)  & 2.29(0.43) \\
\hline
\end{tabular}
\label{tab:gauss}
\end{center}
\end{table}

\end{document}